\documentclass[nohyper,12pt]{JHEP3}
\def\ba{\begin{array}}
\def\ea{\end{array}}
\def\be{\begin{equation}}
\def\ee{\end{equation}}
\def\lbl{\label}
\def \rf{(\ref}

\def\x0{\x_0}
\def\x1{\x_1}

\def\real{{\bf R}}

\def\cd{{\cal {D}}}
\def\cg{{\cal {G}}}
\def\tcg{{\tilde{\cal G}}}

\newtheorem{theorem}{Theorem}
\title{Poisson--Lie T--plurality of three--dimensional conformally invariant sigma models}

\author{L. Hlavat\'y \\ 
Faculty of Nuclear Sciences and Physical Engineering, 
\\ Czech Technical University, 
\\ B\v rehov\'a 7, 115 19 Prague 1, Czech Republic
\\ \email{hlavaty@br.fjfi.cvut.cz} }
\author{L. \v Snobl \\
Faculty of Nuclear Sciences and Physical Engineering, \\
Czech Technical University, \\
B\v rehov\'a 7, 115 19 Prague 1, Czech Republic \\
{\rm and} \\
Centre de Recherches Math\'ematiques, \it Universit\'e de Montr\'eal, \\ 
P.O. Box 6128, Centre-ville Station, Montr\'eal (Qu\'ebec), H3C 3J7 Canada \\
\email{Libor.Snobl@fjfi.cvut.cz} 
}

\abstract{Starting from the classification of 6--dimensional real Drinfeld doubles and  their decomposition into Manin triples 
we construct 3--dimensional Poisson--Lie T--dual or more precisely T--plural sigma models. Of special interest are those 
that are conformally invariant. Examples of models that satisfy vanishing $\beta$--function equations with zero dilaton are presented 
and their duals are calculated. It turns out that for "traceless" dual  algebras they satisfy the $\beta$--function equations as well but usually with rather nontrivial dilaton. We also present explicit examples of several kinds of obstacles and difficulties present in construction of quantum dual models. Such concrete examples might be helpful in further development and improvement of quantum version 
of Poisson-Lie T--duality.
}


\begin{document}

\section{Introduction}
Few years ago, the problem of T-duality of sigma models for the nonabelian groups was solved {on the classical (i.e. non--quantum) level}. As explained in the paper 
\cite{klse:dna} (see also \cite{kli:pltd}), the most apropriate structures for formulation of Poisson--Lie T--dual sigma models are Drinfeld doubles. They are connected Lie 
groups $D$ such that their Lie algebra 
${\cal D}$ 
admits a decomposition into two maximally isotropic subalgebras ${\cal G},{\tilde{\cal G}}$. It means that the dimension of the Drinfeld double must be even and the dimension of the sigma models is equal to the half of the dimension of the double. 

The classification of the two--dimensional models was given in the papers \cite{hlasno:pltdm2dt} and \cite{klva:olr}. Examples of three--dimensional dual models can 
be found e.g. in \cite{vall:su2,sfe:pltd,jare:pltd} but a classification of three--dimensional models is rather complicated. That's why we have restricted to the models that are conformally invariant and do not produce gravitational anomaly. 

In our previous work \cite{snohla:ddoubles} we have performed complete classification of real 6--dimensional Drinfeld doubles including all possible decompositions into the Manin triples and obtained 22 non--isomorphic classes. For all of them we can construct a  variety of models given by three--dimensional subspaces in ${\cal D}$.

In the following sections, we firstly recall the definitions 
of Manin triple and Drinfeld double and briefly explain the construction of dual sigma models. Then we shall present explicit forms of the sigma models that satisfy the vanishing $\beta$--function equations with the zero dilaton field and their duals.

\section{Drinfeld doubles and Manin triples}

The Drinfeld double $D$ is defined as a connected Lie group such that its Lie algebra 
$\cd$ equipped by a symmetric ad--invariant nondegenerate bilinear form 
$\langle\, .,.\rangle $ can be decomposed into a pair of 
subalgebras $\cg$, $\tcg$ maximally isotropic with respect to $\langle\, .,.\rangle $ and $\cd$ as 
a vector space is the direct 
sum of $\cg$ and $\tcg$. This ordered triple of algebras $(\cd,\cg$,$\tcg)$ is called Manin triple. 

One can see that the dimensions of the subalgebras are equal and that bases 
$\{X_i\}, \{\tilde X^i\},\ i=1,2,3$ in the subalgebras can be chosen so that
\be \langle X_i,X_j\rangle =0,\  \langle X_i,\tilde X^j\rangle =\langle 
\tilde X^j,X_i\rangle =\delta_i^j,\  \langle \tilde X^i,\tilde X^j\rangle =0.\lbl{brackets}\ee
Due to the ad-invariance of $\langle\, .,.\rangle $ the algebraic structure of $\cd$ is determined 
by the structure of the maximally isotropic subalgebras because in the basis $\{X_i\}, \{\tilde X^i\}$ 
the Lie bracket is given by 
\[ [X_i,X_j]={f_{ij}}^k X_k,\ [\tilde X^i,\tilde X^j]={\tilde {f^{ij}}_k} \tilde X^k,\]
\be [X_i,\tilde X^j]={f_{ki}}^j \tilde X^k +{\tilde {f^{jk}}_i} X_k. \lbl{liebd}\ee

It is clear that to any Manin triple $({\cal D},{\cal G},{\tilde{\cal G}})$   one can construct the 
dual one by 
interchanging $\cg \leftrightarrow \tcg$, i.e. interchanging the structure coefficients 
$ {f_{ij}}^k \leftrightarrow {\tilde {f^{ij}}_k}$. All properties of Lie algebras 
(the nontrivial being the Jacobi identities) remain to be satisfied. On the other hand for given Drinfeld double more than two Manin triples can exist and there are many examples of that.
Two Drinfeld doubles are isomorphic if they have isomorphic algebraic structure and there is an isomorphism 
transforming one ad-invariant bilinear form to the other. 

Here is the list of nonisomorphic real six--dimensional doubles (see \cite{snohla:ddoubles}) and their decomposition into the Manin triples (for the notation see Appendix \ref{appB}):
\begin{theorem}  \label{veta}
Any 6--dimensional real Drinfeld double belongs just to one of the following 22 classes and allows 
decomposition into all Manin triples listed in the class and their duals $(\cg \leftrightarrow \tcg)$.
If the class contains parameter $a$ or $b$, the Drinfeld doubles with different values of this parameter 
are non--isomorphic. 
\newcounter{Lcount}
\begin{list}{DD\arabic{Lcount}}
{\usecounter{Lcount}
\setlength{\rightmargin}{\leftmargin}}
\item $(9|5|b)$ $\cong $ $(8|5.ii|b)$ $\cong $ $(7_0|5.ii|b)$, $b>0$,
\item $(8|5.i|b)$ $\cong $ $(6_0|5.iii|b)$, $b>0$,
\item $(7_a|7_{1/a}|b)$ $\cong $ $(7_{1/a}|7_a|b)$, $a \geq 1,b \in \real - \{ 0 \},$
\item $(6_a|6_{1/a}.i|b)$ $\cong $ $(6_{1/a}.i|6_a|b)$, $a>1, b \in \real - \{ 0 \},$
\item $(9|1)$,
\item $(8|1)$ $\cong$  $(8|5.iii)$ $\cong$ $(7_0|5.i)$ $\cong$ $(6_0|5.i)$ $\cong$ $(5|2.ii)$,
\item $(7_0|4|b)$  $\cong $ $(4|5.iii|b)$ $\cong$ $(6_0|4.i|-b)$, $b \in \real - \{ 0 \},$  
\item \label{dd33} $(3|3.i|b)$, $b \in \real - \{ 0 \},$
\item $(7_{a}|1)$ $\cong $ $(7_{a}|2.i)$ $\cong $ $(7_{a}|2.ii)$, $a>1$,
\item $(6_a|1)$ $\cong $ $(6_a|2)$ $\cong $ $(6_a|6_{1/a}.ii)$ $\cong $ $(6_a|6_{1/a}.iii)$, $a>1,$
\item \label{dd601} $(6_0|1)$ $\cong$ $(6_0|5.ii)$ $\cong$ $(5|1)$ $\cong$ $(5|2.i)$,
\item $(6_0|2)$ $\cong$ $(6_0|4.ii)$ $\cong$ $(4|1)$ $\cong$ $(4|2.i)$ $\cong$ $(4|2.ii)$,
\item $(3|1)$ $\cong$ $(3|2)$ $\cong$ $(3|3.ii)$ $\cong$ $(3|3.iii)$,
\item $(7_{a}|1)$ $\cong $ $(7_{a}|2.i)$ $\cong $ $(7_{a}|2.ii)$, $0<a<1$,
\item \label{dd701} $(7_0|1)$, 
\item $(7_0|2.i)$,
\item $(7_0|2.ii)$,
\item $(7_{1}|1)$ $\cong $ $(7_{1}|2.i)$ $\cong $ $(7_{1}|2.ii)$,
\item $(2|1)$,
\item $(2|2.i)$,
\item $(2|2.ii)$,
\item $(1|1)$.
\end{list}
\end{theorem}

One can see that for many Drinfeld doubles there are several decompositions into Manin triples. { The (non--isomorphic) Manin triples differing only by the value of the parameter $b$ can be transformed one into another by the rescaling of $\langle \, . , . \rangle$.}

{As we will see in the next section to each Manin triple one can write down a sigma model on the group $G$ (corresponding 
to the 1st subalgebra $\cg$ in the Manin triple) such that its equations of motion can be, at least in the vicinity of the 
group unit, written also 
in the form
\be \langle\partial_\pm l l^{-1},{\cal E}^\pm\rangle=0 \lbl{eqm}\ee
where $l:\real^2 \rightarrow D $  and ${\cal E}^+,{\cal E}^-$ are three--dimensional subspaces of  $\cd$ \cite{klse:dna}. 
Since this form doesn't depend on the choice of Manin triple,  all these models (for fixed 
${\cal E}^\pm$) are equivalent.} {Moreover the scaling of $\langle \,.,. \rangle$ does not change the equations of motion \rf{eqm}) 
and consequently all models corresponding to (non--isomorphic) Drinfeld doubles with different choices of the scaling parameter 
$b$ are equivalent as well.
}
\section{Construction of dual %nonlinear
 sigma models}

Let ${\cal E}^+$ be an $n$--dimensional subspace of $\cd$ and ${\cal E}^-$ its orthogonal complement with respect to $\langle \,.,. \rangle$ such that ${\cal E}^+ + {\cal E}^-=\cd$. This decomposition of $\cd$ defines the equations of motion \rf{eqm}). On the other hand, the decomposition $\cd={\cal G}+{\tilde{\cal G}}$ enables us to write down the equations of motion as those for a sigma model on $G$ (resp. $\tilde G$) {provided ${\cal E}^\pm$ are transversal to ${\cal G},{\tilde{\cal G}}$, i.e. exists a linear invertible map 
\be\label{transv} {\bf E}: {\cal G} \rightarrow {\tilde{\cal G}} \ee
 such that ${\cal E}^+={\rm span} \{ t+ {\bf E}(t), t \in {\cal G} \}$.} Following 
\cite{klse:dna},\cite{kli:pltd}, {we may decompose $l\in D$ in the vicinity of the group unit as  
$$ l= g. \tilde h, \; g \in G, \; \tilde h \in \tilde G $$
(which is always possible on Drinfeld doubles) and by eliminating $\tilde h$ from \rf{eqm}) 
one finds that} the explicit form of the sigma 
 model
equations is given by the Lagrangian 
\be {L}= K_{ij}(g)(g^{-1}\partial_- g)^i(g^{-1}\partial_+ g)^j, \label{lagrangian} \ee
where
\be K(g)=(a(g) + E_0\, b(g))^{-1}E_0\,d(g), \ee
$E_0$ is a constant matrix
and $a(g),b(g),d(g)$ are $n\times n$ submatrices of the adjoint representation of the group $G$ on $\cd$ in the basis $(X_i,\tilde X^j)$
\footnote{ t denotes transposition.}
\be Ad(g)^t  =  \left ( \begin{array}{cc} 
  a(g)&0  \\ b(g)&d(g)  \end{array} \right ), \lbl{adg}\ee
\be a(g)^{-1}=d(g)^t,\ \ \ b(g)^t a(g) = -a(g)^t b(g). \ee
An alternative formulation of the Lagrangian in terms of right--invariant fields that we shall use in the following is
\be {L}=E_{ij}(g)(\partial_- gg^{-1})^i(\partial_+ gg^{-1})^j, \lbl{rflag}\ee
where
\be E(g)=(E_0^{-1}+\Pi(g))^{-1}, \ \ \ \Pi(g)=b(g)a(g)^{-1} = -\Pi(g)^t.\lbl{poiss}\ee
These forms of the Lagrangian guarantee that the sigma model is classically dualizable because if it is rewritten in terms of derivatives of coordinates instead of invariant fields
%given by the Lagrangian
\be {L}= F_{ij}(\phi)\partial_- \phi^i\partial_+ \phi^j, \lbl{sigm}\ee
where $\phi:\real^2\longrightarrow \real^n$ then the covariant tensor field $F$ on the manifold $G$ 
\[ F_{ij}= e_i^a(g)E_{ab}(g)e_j^b(g) \]
satisfies the condition for dualization of the model
\be ( {\cal L}_{ v_{c} } F)_{ij}= \tilde{f}^{ab}_c v_a^mv_b^nF_{im}F_{nj} \lbl{kse}\ee
where $v_a(g)$ are left--invariant vector fields that generate the right action $G$ on itself,
$e^b(g)$ are the dual left--invariant 1--forms,
\[ e_i^a(g) v_a^j(g)=\delta_i^j,\ \ v_a^k(g) e_k^b(g) =\delta_a^b,\]
 and $\tilde f$ are structure constants of a (dual) Lie group.
The covariant tensor  $F$ can be understood as a sum of metric and torsion potential defining the geometric properties of the manifold $G$.
Because (\ref{kse}) is a generalization of condition for a vector field being an isometry, 
the vector fields $v_a(g)$ are called generalized isometries.

The matrix $E_0$ corresponds to the choice of the decomposition $\cd={\cal E}^+ + {\cal E}^-$ and its form can be partially changed by the left translation of G. Namely, if
$ g\mapsto\hat g = hg$ then 
\[ \Pi(\hat g) = b(h)a^{-1}(h) + d(h)\Pi(g)a^{-1}(h) \]
and 
\[ F^{-1}(\hat g) = v^t(g)[\hat E_0^{-1}(h)+\Pi(g)]v(g) \]
where
\be \hat E_0^{-1}(h)=a^t(h)[E_0^{-1}a(h) +b(h)] \ee

{One should also notice that although there are nonisomorphic Manin triples whose commutation relations differ just 
by overall multiplication constant $b$ in all the commutators in the second subalgebra, e.g. the class $(3|3.i|b)$,  
such Manin triples lead to equivalent models. The reason is that such rescaling leads to 
$$ a(g) \rightarrow a(g), \; b(g) \rightarrow b\, b(g), \; d(g) \rightarrow d(g) $$
and consequently to rescaling of the Lagrangian of such model
$$  E_0 \rightarrow \frac{E_0}{b}, \;  L \rightarrow \frac{L}{b}.  $$ 
In the dual model rescaling by $b^{-1}$ occurs. Therefore in the following we always put $$b=1$$ 
and don't mention $b$ explicitly in designations of Manin triples.}

\section{Poisson--Lie T--plurality}
\subsection {Classical level}
{As noted in \cite{klse:dna},\cite{kli:pltd} and investigated explicitly in \cite{unge:pltp}}, the possibility to decompose some Drinfeld doubles into more than two 
Manin triples enables us to construct more than two equivalent sigma models and this property can be called Poisson--Lie T--plurality. 

Let  $\{ T_j,\tilde T^k\},\ j,k\in\{1,...,n\}$ be generators of the Lie algebras of a Manin triple associated with the Lagrangian 
\rf{rflag}) and $\{ U_j,\tilde U^k\}$ be generators of another Manin triple in the same Drinfeld double related by the {$2n\times 2n$ transformation matrix:}

\be \left ( \begin{array}{c} 
  T\\ \tilde T \end{array} \right )
  =  \left ( \begin{array}{cc} 
 P&Q  \\ R&S  \end{array} \right )\left ( \begin{array}{c} U\\ \tilde U \end{array} \right )
\lbl{trsfmat}\ee
The transformed model is then given by the Lagrangian of the same form as  \rf{rflag}) but with $E(g)$ replaced by 
\be E_u(g)=M(N+\Pi_u\,M)^{-1}\lbl{eg} \ee
where
\be M=S^tE_0-Q^t, \ \ N=P^t-R^t E_0, \lbl {mn}\ee
and $\Pi_u$ is calculated by \rf{poiss}) but from the adjoint representation of the group $G_u$ generated by $\{U_j\}$ related to generators of $G$ by \rf{trsfmat}).
Using these formulas we can write down the explicit form of 22 classes of mutually classically equivalent three--dimensional sigma models. 

Examples of Poisson--Lie T--dual models where the matrices in \rf{trsfmat}) are block diagonal resp. block antidiagonal
\[ \left ( \begin{array}{cc} 
 {\bf 1}&0\\ 0&{\bf 1}  \end{array} \right )\ \  {\rm and}\ \  \left ( \begin{array}{cc} 
0&{\bf 1}  \\ {\bf 1}&0  \end{array} \right ) \]
 are sigma models \rf{sigm}) on the doubles $DD20$ and $DD21$ given by tensors
\footnote {We are afraid that the (II,II) model in \cite{jare:pltd} is not selfdual as it does not satisfy the equation \rf{kse}). }
\be F_{ij}(\phi)=(1 + (\phi^3)^2)^{-1}\left(\matrix{ 1 & a\,\phi^3 & \phi^2 \cr - a\,\phi^3 & 1 & - a\,\phi^2\,\phi^3 \cr \phi^2 & a\,\phi^2\,\phi^3 & 
  1 + (\phi^3)^2 +(\phi^2)^2 
 \cr  } \right)
   \ee
\be  \tilde F_{ij}(\phi)=(1 + (\phi^1)^2)^{-1}\left( \matrix{ 1 + ( \phi^1)^2 & 0 & 0 \cr 0 & 1 + 
   ( \phi^1)^2 &  (1 + a)  \,
    \phi^1 \cr 0 & (-1 + a)  \, \phi^1 & 1 \cr  } \right)
\ee 
where $a=1$ for the double $DD20$ and  $a=-1$ for the double $DD21$.
Both these models have nonzero and nonconstant torsion and scalar curvature.

The elements of $G$ and $\tilde G$ are  parametrized as
$$ g=e^{\phi^1X_1}e^{\phi^2X_2}e^{\phi^3X_3}, \; { \tilde g=
e^{\tilde \phi_1 \tilde X^1}e^{\tilde \phi_2 \tilde X^2}e^{\tilde \phi_3 \tilde X^3}.} $$ This convention will be used further as well. In order not to overburden the reader with too complicated notation we shall always assume that $\phi$s denote the coordinates 
on the group just under consideration. Therefore $\phi$'s in different models on the same Drinfeld double denote 
coordinates on different groups.

\subsection {Quantum level -- conformal invariance}

The equivalence on quantum level is  more complicated \cite{alkltse:qpl,tvu:pltdpi}. 
In quantum theory the duality or plurality transformation must be supplemented by a correction that comes from integrating out 
the fields on the dual group $\tilde G$ in the path integral formulation. In some cases {it} can be absorbed {at the 1-loop level} into the transformation of the dilaton field $\Phi$. The dilaton field can be understood as an additional function on $G$ that defines the nonlinear sigma model and couples to scalar curvature of the manifold. The transformation of the tensor $F$ then must be acompanied by the transformation of the dilaton \cite{unge:pltp}
\be  \Phi_u= \Phi+{\rm ln\ Det} (N + \Pi_u M) - {\rm ln\ Det} ({\bf 1} + \Pi E_0)+{\rm ln\ Det}\,a_u-{\rm ln\ Det} \, a  \lbl{dualdil}\ee
where $\Pi_u, a_u$ are calculated as in \rf{adg}),\rf{poiss}) but from the adjoint representation of the group $G_u$.

The condition that the theory is ultraviolate finite is equivalent to the conformal invariance of the model. That
is expressed by the {vanishing} $\beta$--function equations.\footnote{To be more precise, we tacitly assume that there are 
other 23 dimensions, in which spectator fields live, as required by criticality of bosonic strings. Thus the term
$D-26$ that should appear in (\ref{bt3}) vanishes but we suppose that 
these 23 extra dimensions form a space which is flat and completely decouples from the group $G$. It means that the spacetime is 
assumed to be $G \times \real^{23}$ (resp. $G \times \real^{q} \times T^{23-q}$)
and coordinates on the flat component and corresponding components of tensors etc. don't appear in the equations (\ref{bt1}--\ref{bt3}).}
At the one loop level these equations read
\begin{eqnarray}
\label{bt1} 0 & = & R_{ij}-\bigtriangledown_i\bigtriangledown_j\Phi- \frac{1}{4}H_{imn}H_j^{mn} \\ 
 \label{bt2} 0 & = & \bigtriangledown^k\Phi H_{kij}+\bigtriangledown^k H_{kij}  \\
 \label{bt3} 0 & = & R-2\bigtriangledown_k\bigtriangledown^k\Phi- \bigtriangledown_k\Phi\bigtriangledown^k\Phi- \frac{1}{12}H_{kmn}H^{kmn}
 \end{eqnarray}
where covariant derivatives $\bigtriangledown_k$, Ricci tensor $R_{ij}$ and scalar curvature $R$ are calculated from the metric
\be G_{ij}=\frac{1}{2}(F_{ij}+F_{ji}) \ee
which is also used for lowering and raising indices and 
\be H_{ijk}=\partial_i B_{jk}+\partial_j B_{ki}+\partial_k B_{ij} \lbl{torsion} \ee
where
\be B_{ij}=\frac{1}{2}(F_{ij}-F_{ji}). \lbl{torpot}\ee

\section{Conformally invariant sigma models and their duals}
To choose from the huge variety of three dimensional sigma models we looked for {those} that satisfy the 
 1--loop vanishing $\beta$--function equations. 
 
{As opposed to} the metric and torsion, the Poisson--Lie T--duality gives no hint what the possible dilaton fields are. That's why we have chosen the simplest possibility and tested which dualizable models satisfy the equations (\ref{bt1}--\ref{bt3})  for the dilaton field
{equal to zero}. For technical reasons we have restricted ourselves to the models on solvable Drinfeld doubles 
and the diagonal input matrices
\be E_0=\left(\matrix{ p&0&0\cr 0&q&0 \cr 0&0&r
 \cr  } \right)
   \ee
In addition to the already known abelian model we have found three models with zero dilaton namely those corresponding to the Manin triples $(6_0|1)$, $(7_0|1) $ and $(3|3.i) $. All these models are flat and torsionless but their duals do not share these properties.

\subsection{Selfdual double $DD$\ref{dd33}} 
The first example of a model that satisfies the {vanishing} $\beta$--function equations \rf{bt1})--\rf{bt3}) with $\Phi\equiv 0$ is  that produced by the (only possible) decomposition of the double $DD$\ref{dd33} into $(3|3.i)$ and
 \footnote{ $\pm$ signs represent the same sign in all the 
following expressions.}
$$ p= \pm 1, \; q= r=\pm k. $$
This gives
\footnote{Recall that 
$F_{ij}(0)=E_{ij}(e)=(E_0)_{ij}$.}
\be F_{ij}=\pm k(e^{4\phi^1}+2k\,\rho^2)^{-1}\left(\matrix{ k^{-1}e^{4\phi^1} & -\rho & -\rho \cr \rho & \frac{1}{2}(1+e^{4\phi^1})+k\,\rho^2 & \frac{1}{2}(1-e^{4\phi^1})-k\,\rho^2 \cr \rho & \frac{1}{2}(1-e^{4\phi^1})-k\,\rho^2 & \frac{1}{2}(1+e^{4\phi^1})+k\,\rho^2 
 \cr  } \right)
   \ee
where 
\[ \pm\rho=\phi^2+\phi^3. \]   
In spite of the fact that the tensor field $F$ is not symmetric the torsion $H$ calculated by \rf{torsion}) and \rf{torpot}) vanishes as well as the Riemann tensor.

This model is classically selfdual, i.e. the tensor field $\tilde F_{ij}$ is obtained from $F_{ij}$ only by $k\rightarrow 1/k$, but it is well known fact that models defined for Manin triples whose structure coefficients have nonzero trace $\tilde f_{ij}^j$ (which is the case of the Bianchi algebra {\bf 3}) are anomalous on the quantum level. It means that the effective action has a gravitational anomaly that cannot be absorbed into the transformation of the dilaton. If one nevertheless tries to construct the dual dilaton using (\ref{dualdil}), one finds that it depends
on the coordinates on both subgroups. Since the dilaton must not depend on the subgroup integrated out in path integral, 
 it is not well defined in this case. Such problem was encountered already in models constructed in \cite{unge:pltp} and its solution is still unknown, presumably it will require some modification of (\ref{dualdil}).

\subsection{T--Duality in $DD$\ref{dd701}}
The second example of a sigma model that satisfies the {vanishing} $\beta$--function equations with $\Phi\equiv 0$ is the model with the 
metric 
\be F_{ij}= \left(\matrix{ p&0 &p\,\phi^2\cr 0 & p & -p\,\phi^1\cr p\phi^2& -p\phi^1 & r+p\,(\phi^1)^2 +p\,(\phi^2)^2
 \cr  } \right)
\ee
obtained from the decomposition of the Drinfeld double $DD$\ref{dd701} into $(7_0|1)$. The tensor field of the dual model corresponding to $(1|7_0)$ is 
\be \tilde F_{ij}= \left[p\,r  +(\phi^1)^2)+(\phi^2)^2\right]^{-1}
\left(\matrix{ 
r+\frac{1}{p}(\phi^1)^2 & \frac{1}{p}\phi^1\phi^2 
& -\phi^2 
\cr \frac{1}{p}\phi^1\phi^2  & r+\frac{1}{p}(\phi^2)^2  & \phi^1
\cr \phi^2 & -\phi^1 & p
\cr  } \right)
\lbl{df17}\ee
The dual dilaton field 
calculated from \rf{dualdil}) is 
\be \tilde\Phi(\phi)={\rm ln}\left(-p\,\left(r\,p+(\phi^1)^2+(\phi^2)^2\right)\right)
         \ee
and together with the tensor field \rf{df17}) they satisfy the {vanishing} $\beta$--function equations so that the duality is preserved also on the quantum level.

Note that in the case of  $p\,r <0$  both the tensor and dilaton fields 
are singular for \be (\phi^1)^2+(\phi^2)^2=-p\,r. \lbl{sinm17}\ee
At these points curvatures are infinite,  even the scalar one
$$ R =  2p\frac{5p\,r-2(\phi^1)^2-2(\phi^2)^2}{{
(p\,r+(\phi^1)^2+(\phi^2)^2)}^2}, $$
therefore there is a genuine singularity on the hypersurface \rf{sinm17}) in the target manifold.

In the $\beta$--function equations the singularities of curvature, torsion and dilaton cancel provided one takes limits in 
(\ref{bt1}--\ref{bt3}) .

By using transformation (\ref{trsfmat}) transforming $(7_0|1)$ 
into isomorphic Manin triples, i.e. the same Manin triple $(7_0|1)$ but immersed 
in a different ways into $DD$\ref{dd701}, one may also get models with nondiagonal matrix $E_0$ on the same group $G_{{\bf 7_0}}$. 
By explicit calculation one finds the models with
\be E_0 = \left(\matrix{ e_{11} & e_{12} &e_{13} \cr -e_{12} & e_{11} & e_{23} \cr e_{31} & e_{32} & e_{33}
 \cr  } \right),
\ee
where 
$$ e_{33} = \frac{4 r e_{11}+e_{32}^2+2 e_{13} e_{31}+e_{31}^2+e_{13}^2+e_{23}^2+2 e_{23} e_{32}}{4 e_{11}} $$
and $e_{11},e_{12},e_{13},e_{23},e_{31},e_{32}$ 
are arbitrary constants such that ${E_0}^{-1}$ exists, and nonzero but constant dilaton\footnote{ 
 Note that consequently all models (\ref{df17}) with different choices of $p$ are equivalent. Different choices of $r$ 
give models which are not mutually Poisson--Lie T--plural, but $r$ can be absorbed into overall factor in the Lagrangian using 
the fact that Manin triple $(7_0|1)$ is isomorphic to Manin triples $(7_0|1|b)$ where $\langle \,.,. \rangle$ is rescaled 
by $b$.}. These models have 
nonvanishing torsion potential, but vanishing torsion and curvature. Therefore they differ from the original one only 
by closed, i.e. locally exact, form in Lagrangian and constant dilaton shift and are consequently  up to global issues equivalent 
to the original one (both on the classical and for most applications also on the quantum level). Similarly, one may also consider 
corresponding models on the dual group $G_{{\bf 1}}$ but again it seems that nothing qualitatively new emerges, only the 
expressions get 
rather complicated and not suitable for presentation here, e.g. the dilaton now  up to constant shift reads 
$$ \ln(-e_{11}^2 e_{33}+e_{23} e_{32} e_{11}+e_{11} e_{23} \phi^1-e_{11}  e_{32} \phi^1 -e_{11} 
(\phi^1)^2-e_{33} e_{12}^2+$$ 
$$+e_{13} e_{12} e_{32}+
e_{12} e_{13} \phi^1+e_{12}  e_{32} \phi^2
-e_{31} e_{12} e_{23}+e_{31} e_{12} \phi^1+$$ 
$$+e_{31} e_{11} e_{13}+e_{31} e_{11} \phi^2+ e_{23} e_{12} \phi^2- e_{11} e_{13} \phi^2-e_{11} (\phi^2)^2). $$

\subsection{Poisson--Lie T--plurality in $DD$\ref{dd601} }
This is the most interesting example as this double can be decomposed into more than two dual Manin triples. 
It was investigated with a spectator dependent $E_0$ in \cite{unge:pltp}. Even if no spectator dependence is assumed then there is a sigma model with zero dilaton and diagonal $E_0$, namely that corresponding to the Manin triple $(6_0|1)$.

None of the models given by other decompositions of this Drinfeld double and diagonal $E_0$ satisfy the $\beta$--function equations with $\Phi\equiv 0$. Nevertheless, we can find sigma models with nonzero dilaton fields corresponding to the other decompositions using transformations between different Manin triples and check the $\beta$--function equations. These models have nontrivial metrics, torsions and dilatons.

\subsubsection{Decompositions $(6_0|1)$ and $(1|6_0)$}
As mentioned, the simplest model comes from the semiabelian decomposition $(6_0|1)$ that produces sigma model with the metric
\be F_{ij}= \left(\matrix{ p&0 &p\,\phi^2\cr 0 & -p & -p\,\phi^1\cr p\phi^2& -p\phi^1 & r-p\,(\phi^1)^2 +p\,(\phi^2)^2
 \cr  } \right)
\label{model601diag}\ee
that is flat (and torsionless) so that the model satisfies the $\beta$--function equations with the vanishig dilaton field. The tensor field of the dual model is 
\be \tilde F_{ij}= \left[p\,r  -(\phi^1)^2+(\phi^2)^2\right]^{-1}
\left(\matrix{ 
r-\frac{1}{p}(\phi^1)^2 & \frac{1}{p}\phi^1\phi^2 
& \phi^2 
\cr \frac{1}{p}\phi^1\phi^2  & -r-\frac{1}{p}(\phi^2)^2  & -\phi^1
\cr -\phi^2 & \phi^1 & p
\cr  } \right)
\lbl{df16}\ee
This model is neither flat nor torsionless, nevertheless the $\beta$--function equations \rf{bt1})--\rf{bt3}) are satisfied for the dual dilaton field 
\be \tilde\Phi(\phi)={\rm ln}\left(p\left(r\,p-(\phi^1)^2+(\phi^2)^2\right)\right)
         \ee
calculated from \rf{dualdil}). 
Note that, like in the $DD$\ref{dd701} case, both the tensor and dilaton fields 
are singular for \be (\phi^1)^2=p\,r+(\phi^2)^2. \lbl{sinm}\ee
At these points curvatures are infinite,  even the scalar one
$$ R = -2 p\frac{5p\,r+2(\phi^1)^2-2(\phi^2)^2}{
(p\,r-(\phi^1)^2+(\phi^2)^2)^2},$$
therefore there is a genuine singularity on the hypersurface \rf{sinm}) in the target manifold.
 In the $\beta$--function equations the singularities of curvature, torsion and dilaton cancel.

{By using transformation (\ref{trsfmat}) transforming $(6_0|1)$ 
into isomorphic Manin triples, similarly as in $DD$\ref{dd701} case one may again 
get models with nondiagonal matrix $E_0$ on the same group $G_{{\bf 6_0}}$. 
By explicit calculation one finds 
models with
\be E_0 = \left(\matrix{ e_{11} & e_{12} &e_{13} \cr -e_{12} & -e_{11} & e_{23} \cr e_{31} & e_{32} & e_{33}
 \cr  } \right),
\ee
where 
$$ e_{33} =  \frac{4 r e_{22} - e_{31}^2  + e_{32}^2  + 2 e_{23} e_{32} - 2 e_{31} e_{13} + e_{23}^2  - e_{13}^2 
}{4 e_{22}} $$
and $e_{11},e_{12},e_{13},e_{23},e_{31},e_{32}$ are arbitrary constants such that ${E_0}^{-1}$ exists, and 
constant dilaton.\footnote{ Note that consequently all models (\ref{model601diag}) with different choices of $p$ are equivalent. 
 Different choices of $r$ 
give models which are not mutually Poisson--Lie T--plural, but using the fact that Manin triple $(6_0|1)$ is isomorphic to Manin 
triples $(6_0|1|b)$ where $\langle \,.,. \rangle$ is rescaled by $b$ one may extract $r$ as an overall factor in  the Lagrangian 
and in this sense all models (\ref{model601diag}) and all their Poisson--Lie T--plurals are equivalent.}
These models again, as in $DD$\ref{dd701} case, have nonvanishing torsion potential, but vanishing torsion and curvature, and, as in in $DD$\ref{dd701} case, are up to global issues equivalent to the original one. 
When one considers the corresponding models on the dual group $G_{{\bf 1}}$ it again seems that nothing really  new and interesting
emerges, only the expressions get rather complicated, e.g. the dilaton now  reads (up to constant shift)
$$ \ln(e_{32} e_{12} e_{13}-e_{23} e_{12} e_{31}-\phi^2 e_{32} e_{12}-\phi^2 e_{23} e_{12}+\phi^1 e_{13} e_{12}+$$
$$+ e_{31} e_{12}\phi^1-e_{11} (\phi^1)^2-e_{12}^2 e_{33}+e_{33} e_{11}^2+e_{11} (\phi^2)^2+e_{11}e_{23}\phi^1 +$$
$$+e_{11} \phi^2 e_{31}-e_{31} e_{11} e_{13}-e_{11} \phi^2 e_{13}-e_{11} \phi^1 e_{32}+e_{32} e_{23} e_{11}).  $$}

\subsubsection{Decomposition $( 5|2.i)$ }
The general form of the transformation matrix \rf{trsfmat}) from $(6_0|1)$ to $( 5|2.i)$ is rather complicated and contains eight free real parameters. Nevertheless, using \rf{trsfmat}) and\rf{mn}) one can calculate the sigma model corresponding to the decomposition $( 5|2.i)$ and equivalent to the previous two. The model is given by the tensor field
\be  F_{ij}(\phi)= 
\left(\matrix{ 
r+\gamma\,Y( a,b)Y( c,d)
& Y( a,d)&
\gamma\,Y( a(\beta-1),d(\beta+1))
\cr Y( c,b) &\gamma^{-1} & \beta-1
\cr \gamma\,Y( c(1+\beta),b(\beta-1)) &\beta+1&\gamma (\beta^2-1)
\cr  } \right)
\lbl{df52}\ee
where 
$$ Y( a,b)=a\,e^{\phi^1}+b\,e^{-\phi^1} $$
and $a,b,c,d,\beta,\gamma$ are real parameters coming from the transformation matrix and $p,r$.
Dilaton field is (rather surprisingly) constant because this model is again flat and torsionless.
This is another example of a model with constant dilaton field but nondiagonal matrix $\tilde E_0=MN^{-1}=F(0)$.
This is also the only case we know where the metric is flat but the necessary condition for dualizability (\ref{kse})
is satisfied non--trivially -- neither left nor right hand side of it vanishes {\it per se}, 
i.e. we consider an algebra of genuine generalized isometries even for trivial, flat metric.

One can calculate the model corresponding to $(2.i|5)$ as well but as the Bianchi algebra {\bf 5} is not traceless it does not satisfy the $\beta$--function equation \rf{bt3}).

\subsubsection{Decomposition $(5.ii|6_0)$}
The general form of the transformation matrix \rf{trsfmat}) from $(6_0|1)$ to $(5.ii|6_0)$ contains seven free real parameters and the general form of the sigma model is so complicated that it is hardly of any use and nearly impossible to display. We shall present here its very special form corresponding to the symmetric matrix 
$$\tilde E_0 = \left(\matrix{ 
r& -r&0
\cr -r &-3r &0
\cr 0&0&{1}/{r}
\cr  } \right). $$
The model is given by the tensor field 
\be F_{ij}= (-\lambda  + 2\,{\lambda }^2 + 
  2\,\varrho  - 2\,\lambda \,\varrho)^{-1}\left(\matrix{r\,\alpha&r\,\beta &\gamma\cr r\,\beta & r\,\epsilon & \delta\cr -\gamma& -\delta & \lambda / r
 \cr  } \right)
\lbl{df560}\ee
where
\[ \rho= 2e^{\phi^1} \]
\[ \lambda = e^{(\phi^1 + \phi^2)} \]   
   
\[ \alpha=
     \left( \lambda - \varrho  \right) \,
     \left( -2 + \varrho \,\lambda  - 
       {\lambda }^2 \right)  \]
\[ \beta=-
     \left( -\varrho  + {\varrho }^2 + 
       2\,\lambda  - \varrho \,\lambda  - 
       \varrho \,{\lambda }^2 + {\lambda }^3
       \right) \]
\[ \gamma = \lambda \,
   \left( 1 - \varrho  + \lambda  \right) \]
\[ \delta = \varrho  + \lambda  - 
   2\,\varrho \,\lambda  + {\lambda }^2 \]
\[ \epsilon = -
      { \left( {\varrho }^2 + 
         2\,{\lambda }^2 - 
         2\,\varrho \,{\lambda }^2 + 
         {\lambda }^4 \right) }{\lambda }^{-1} \]
The dilaton for this model is of the form
\be \Phi= \ln \,\left(\frac{1}{2}\,(   2\,e^{\phi_1 + \phi_2}
 - 4\,e^{\phi_1} + {4}\,{e^{-\phi_2}} -1
 )\right)\ee 
and together with the tensor field \rf{df560}) satisfy the {vanishing} $\beta$--function equations so that the duality on the quantum level is preserved.

Again the metric, its curvature and the dilaton are singular for $\phi_1, \phi_2$ satisfying
$$ 2\,e^{\phi_1 + \phi_2} - 4\,e^{\phi_1} + {4}\,{e^{-\phi_2}} = 1. $$
and in the $\beta$--function equations the singularities cancel against each other.

In principle one can also calculate the model corresponding to $(6_0|5.ii)$ but cannot expect that it satisfies the $\beta$--function equations  as the Bianchi algebra {\bf 5} is not traceless.
\subsubsection{Note on $(5|1)$}
The remaining decompositions of the Drinfeld double $DD$\ref{dd601} are the Manin triples $(5|1)$ and $(1|5)$. We must discard $(1|5)$ for the same reason as $(2.i|5)$ and $(6_0|5.ii)$ but one naturally expects that the decomposition $(5|1)$ may produce another conformally invariant sigma model. Unfortunately this is not true. The reason is that $\Pi_u=0$ (it holds for any $(X|1)$) so that $E_u(g)=MN^{-1}$ and the matrix $N$ calculated by \rf{mn}) is singular for any transformation matrix from $(6_0|1)$ to $(5|1)$ (in spite of the fact that they contain again several free parameters).
More generally speaking in this case the transversality condition (\ref{transv}) on ${\cal E}^\pm$ 
is not satisfied for any values of parameters $p,r$ and any choice of Manin triple isomorphic to $(5|1)$.

{The cases that the transversality condition 
is not satisfied occur for the Manin triples  $(5|2.i)$ and $(5ii|6_0)$ as well but only for special choices of parameters in the transformation matrices \rf{trsfmat}). It is interesting that even in these cases the metrics of the models are not everywhere singular.}

\section{Conclusions}

\subsection{Difficulties encountered on the classical level}

As we have seen, in construction of dual models one may encounter several difficulties. Firstly, on the classical level 
one might naively assume that one may go to any Manin triple in the considered Drinfeld double and construct models on the 
corresponding groups. This is true only in the generic case, for special value of parameters in $E_0$ the transversality 
condition (\ref{transv}) might not be satisfied for some decompositions into Manin triple (as seen in the case of 
Drinfeld double $DD$\ref{dd601} and Manin triple $(5|1)$). Unfortunately, these cases might just be the ones satisfying 
some other requirements on the models (e.g. conformal invariance). Also if one considers the case with nontrivial spectators 
then the matrix $E_0$ depends on them and the duality may break down for some specific values of the spectator fields, e.g.
on some hyperplane in the 4--dimensional spacetime. { 
We don't know any way to circumvent
such obstacle if it occurs.}

It follows that also the notion of modular space of Drinfeld double 
defined in \cite{klse:dna} and investigated e.g. in 
\cite{sno:modsp},\cite{sno:msldd} 
should be made more precise. We realize that whether it is possible to 
construct a model on  certain Manin triple of Drinfeld double or not 
depends  via the transversality condition on the subspace ${\cal E}^+$ or equivalently 
on the matrix $E_0$. It means that the modular space 
as the set of all equivalent Poisson--Lie T--dual models depends not only 
on the Drinfeld double, but also on the choice of the subspace\footnote{More precisely it depends on the equivalence class 
of these subspaces ${\cal E}^+$ with respect to $\langle\, .,.\rangle $--preserving automorphisms of $\cd$; the equivalent subspaces 
 give rise to isomorphic modular spaces that can be identified by change of basis in $\cd$.} 
${\cal E}^+$; for different 
${\cal E}^+$ the modular spaces may be nonisomorphic. The results 
obtained before in \cite{sno:modsp},{sno:msldd}, where modular space was identified 
with the set of all decomposition of Drinfeld double into Manin triples, 
are therefore valid only for generic ${\cal E}^+$, i.e. such that the transversality 
condition (\ref{transv}) is satisfied for all possible Manin triples in the Drinfeld 
double. For certain special matrices $E_0$ transformation to some Manin 
triples might be ruled out and the modular space is then only a subset of the 
generic one.  

Also we have obtained singular metrics as duals to ordinary flat metric. 
This might look rather surprising because the manifolds considered are in fact Lie groups but 
it gives an explicit example of local nature of Poisson--Lie T--duality (and also of its special version, 
the so--called nonabelian or semiabelian T--duality, where $\tcg$ is abelian). From construction the duality 
transformation is guaranteed to exists only in some neighborhood of group unit element and encountering singularity signals 
that one has applied the present formalism of T--duality beyond its domain of applicability. 
Nevertheless, the resulting models still seem to be interesting examples of singular conformally invariant string backgrounds 
and might become well--defined duals in some future reformulation of Poisson--Lie T--duality. They might be 
interpreted as spacetimes with brane--like objects.

\subsection{Dilaton and quantum Poisson--Lie T--duality}
In the quantum case the dilaton field is known to cause problems. Although its transformation was constructed in \cite{unge:pltp}
 so that 
the 1--loop conformal invariance, i.e. vanishing $\beta$--function equations, is satisfied in the dual model,
the dilaton might not be a well defined object in the dual theory at all. Namely it might dependent 
on the coordinates on the dual group presumably integrated out, as seen already in \cite{unge:pltp}. 
This occurs here for $(3|3.i)$ (which is in any case problematic because of nonvanishing trace).
Also from a mathematical point of view one tends to believe that if the Poisson--Lie T--plurality 
is to be well defined also on the 1--loop quantum level, there should exist a better description of the dilaton, 
namely as an object on the whole double not depending on the concrete choice of Manin triples. The dilatons in respective 
dual theories shall then follow from it.
 
Also one may observe that in each of the classes of Poisson--Lie T--plural models on $DD$\ref{dd601} and $DD$\ref{dd701} 
we have one flat model with zero torsion potential and zero dilaton. 
(The metrics of $(6_0|1)$ and $(7_0|1)$ become constant and diagonal for coordinates $\chi$ obtained from the group element 
parametrization $g=e^{\chi_3X_3}e^{\chi_2X_2}e^{\chi_1X_1}$). Since for the considerations 
in string theory the group plays only an auxiliary role, the important data are the target manifold and its metric 
(together with torsion potential and dilaton), for practical purposes all these models appear to be 
 equivalent and are also equivalent to the model on $(1|1)$ with Minkowski metric. All equivalence is of course only local, 
as was already stressed, global issues are not at the present level of understanding covered by Poisson--Lie  T-duality at all. 

\subsection{Future prospects}
After sheding some light on complications occuring in study of 1--loop conformally invariant 
Poisson--Lie T--dual models we would like to express our opinion that in spite of the current, still very limited, 
understanding of quantum properties of Poisson--Lie T--duality it might already give us some practical results, 
e.g. can be used to generate rather nontrivial examples of conformally invariant 
string backgrounds. {We hope that explicit examples presented here can be used as nontrivial ``guinea pigs''
in future attempts to improve quantum version of Poisson-Lie T--duality and that the improved version 
of it might avoid some of the pitfalls encountered in the current investigation.}

\acknowledgments{ Support of the Ministry of Education of Czech Republic
under the research plan MSM210000018 is gratefully acknowledged. L. \v Snobl thanks 
Centre de Recherches Math\'ematiques for the support of his postdoctoral stay at Universit\'e de Montr\'eal.
We are also grateful to Rikard von Unge for several interesting and helpful discussions and for his comments 
on early versions of the manuscript.}

\appendix

\section{Bianchi algebras}\label{appA}
It is known that any 3--dimensional real Lie algebra can be brought to one of 11 forms by a change of basis.
These forms represent non--isomorphic Lie algebras and are conventionally known as Bianchi algebras.  
They are denoted by  ${\bf 1},\ldots,{\bf 5}$, ${\bf 6_a}$,${\bf 6_0}$,
${\bf 7_a}$,${\bf 7_0}$,${\bf 8}$,${\bf 9}$
 (see e.g. \cite {Landau}, in literature often uppercase 
roman numbers are used instead of arabic ones).The corresponding Lie groups we denote e.g. by $G_{{\bf 1}}$.
 The list of Bianchi algebras is given in decreasing order 
starting from simple algebras.

\begin{description}
\item[${\bf 9:}$] $[X_1,X_2]=X_3, \; [X_2,X_3] = X_1, \; [X_3,X_1] = X_2, $ (i.e. $so(3)$)
\item[${\bf 8:}$] $[X_1,X_2]=-X_3, \; [X_2,X_3] = X_1, \; [X_3,X_1] = X_2, $ (i.e. $sl(2,{\real})$)
\item[${\bf 7_a:}$] $[X_1,X_2]=-a X_2+X_3, \; [X_2,X_3] = 0, \; [X_3,X_1] = X_2+ a X_3, 
\; a >0, $
\item[${\bf 7_0:}$] $[X_1,X_2]=0, \; [X_2,X_3] = X_1, \; [X_3,X_1] = X_2, $
\item[${\bf 6_a:}$] $[X_1,X_2]=-a X_2-X_3, \; [X_2,X_3] = 0, \; [X_3,X_1] = X_2+ a X_3, 
\; a >0, \, a \neq 1 , $
\item[${\bf 6_0:}$] $[X_1,X_2]=0, \; [X_2,X_3] = X_1, \; [X_3,X_1] = -  X_2, $
\item[${\bf 5:}$] $[X_1,X_2]=-X_2, \; [X_2,X_3] = 0, \; [X_3,X_1] = X_3, $
\item[${\bf 4:}$] $[X_1,X_2]=-X_2+X_3, \; [X_2,X_3] = 0, \; [X_3,X_1] = X_3, $
\item[${\bf 3:}$] $[X_1,X_2]=-X_2-X_3, \; [X_2,X_3] = 0, \; [X_3,X_1] = X_2+X_3, $
\item[${\bf 2:}$] $[X_1,X_2]=0, \; [X_2,X_3] = X_1, \; [X_3,X_1] = 0,$
\item[${\bf 1:}$] $[X_1,X_2]=0, \; [X_2,X_3] = 0, \; [X_3,X_1] = 0 , $
\end{description}

\section{List of Manin triples}\label{appB}
We present a list of Manin triples based on \cite{hlasno:cfn6mt}. 
The label of each Manin triple, e.g. 
${\bf  ( 8|5.ii|b)  }$, indicates the structure of the first subalgebra  $\cg$, 
e.g. Bianchi algebra ${\bf 8}$, the structure of the second subalgebra  $\tcg$, e.g. Bianchi algebra  
${\bf 5}$;
roman numbers $i$, $ii$ etc. (if present) distinguish between several possible 
pairings $\langle \, . , . \rangle$ of the subalgebras $\cg,\tcg$ and the parameter $b$ indicates
the Manin triples differing by the rescaling of $\langle \, . , . \rangle$
(if such Manin triples are not isomorphic).

The Lie structures of the subalgebras $\cg$ and $\tcg$ are written out in mutually dual bases 
$(X_1,X_2,X_3)$ and $(\tilde{X}^1,\tilde{X}^2,\tilde{X}^3)$ where a transformation 
was used to bring $\cg$ to the standard Bianchi form (therefore its structure is given in 
Appendix \ref{appA} and not listed here). Because of (\ref{liebd})
this information specifies the Manin triple completely.

The dual Manin triples ($\cd,\tcg, \cg$) are not written explicitly
 but can be easily obtained by $X_j\leftrightarrow\tilde X^j$.

\begin{enumerate}
\item Manin triples with the first subalgebra $ \cg= {\bf 9}$:
\begin{description}
\item[${\bf (9|1 ) } $] :
 $ [\tilde{X}^1,\tilde{X}^2]= 0 ,  \, 
[\tilde{X}^2,\tilde{X}^3] = 0 , \; 
[\tilde{X}^3,\tilde{X}^1] = 0. $ 
\item[${\bf (9|5|b )}$] :
  $ [\tilde{X}^1,\tilde{X}^2]=- b \tilde{X}^2, \, 
[\tilde{X}^2,\tilde{X}^3] = 0 , \; 
[\tilde{X}^3,\tilde{X}^1] = b \tilde{X}^3, \; b > 0  . $ 
\end{description}

\vskip2mm  \item Manin triples with the first subalgebra $ \cg= {\bf 8}$:
\begin{description} 
   \item[${\bf ( 8|1) } $] : 
$[\tilde{X}^1,\tilde{X}^2]= 0 ,  \, 
[\tilde{X}^2,\tilde{X}^3] = 0 , \; 
[\tilde{X}^3,\tilde{X}^1] = 0. $ 
   \item[${\bf ( 8|5.i|b) } $] : 
$ [\tilde{X}^1,\tilde{X}^2]=- b \tilde{X}^2, \, 
[\tilde{X}^2,\tilde{X}^3] = 0 , \; 
[\tilde{X}^3,\tilde{X}^1] = b \tilde{X}^3, \; b >0 . $ 
   \item[${\bf ( 8|5.ii|b) } $] : 
$ [\tilde{X}^1,\tilde{X}^2]=0, \, 
[\tilde{X}^2,\tilde{X}^3] = b  \tilde{X}^2 , \; 
[\tilde{X}^3,\tilde{X}^1] = - b \tilde{X}^1, \; b >0 . $ 
   \item[${\bf ( 8|5.iii) } $] : 
$ [\tilde{X}^1,\tilde{X}^2]= \tilde{X}^2, \, 
[\tilde{X}^2,\tilde{X}^3] =   \tilde{X}^2  , \; 
[\tilde{X}^3,\tilde{X}^1] = - ( \tilde{X}^1+\tilde{X}^3 ) . $
\end{description}

\vskip2mm  \item Manin triples with the first subalgebra $ \cg= {\bf 7_{a}}$:
\begin{description} 
   \item[${\bf ( 7_a|1) } $] : 
$  [\tilde{X}^1,\tilde{X}^2]= 0, \, 
[\tilde{X}^2,\tilde{X}^3] = 0 , \; 
[\tilde{X}^3,\tilde{X}^1] = 0. $ 
   \item[${\bf ( 7_a|2.i) } $] : 
$ [\tilde{X}^1,\tilde{X}^2]=0, \, 
[\tilde{X}^2,\tilde{X}^3] = \tilde{X}^1 , \; 
[\tilde{X}^3,\tilde{X}^1] = 0 .$ 
   \item[${\bf ( 7_a|2.ii) } $] : 
$ [\tilde{X}^1,\tilde{X}^2]=0, \, 
[\tilde{X}^2,\tilde{X}^3] = - \tilde{X}^1 , \; 
[\tilde{X}^3,\tilde{X}^1] = 0 . $
   \item[${\bf ( 7_a|7_{1/a}|b) } $] :
$ [\tilde{X}^1,\tilde{X}^2]= b ( -  \frac{1}{a} \tilde{X}^2 + \tilde{X}^3), \, 
[\tilde{X}^2,\tilde{X}^3] = 0,$ \\ 
$[\tilde{X}^3,\tilde{X}^1] = b (\tilde{X}^2+ \frac{1}{a} \tilde{X}^3), 
\; b \in { \real} - \{ 0 \} . $ 
\end{description}

\vskip2mm  \item Manin triples with the first subalgebra $ \cg= {\bf 7_0}$:
\begin{description} 
   \item[${\bf ( 7_0|1) } $] : 
$ [\tilde{X}^1,\tilde{X}^2]= 0, \, 
[\tilde{X}^2,\tilde{X}^3] = 0 , \; 
[\tilde{X}^3,\tilde{X}^1] = 0. $ 
   \item[${\bf ( 7_0|2.i) } $] : 
$ [\tilde{X}^1,\tilde{X}^2]= \tilde{X}^3, \, 
[\tilde{X}^2,\tilde{X}^3] = 0 , \; 
[\tilde{X}^3,\tilde{X}^1] = 0 .$ 
   \item[${\bf ( 7_0|2.ii) } $] : 
$ [\tilde{X}^1,\tilde{X}^2]= - \tilde{X}^3, \, 
[\tilde{X}^2,\tilde{X}^3] = 0 , \; 
[\tilde{X}^3,\tilde{X}^1] = 0 . $
   \item[${\bf ( 7_0|4|b) } $] : 
$ [\tilde{X}^1,\tilde{X}^2]= b ( - \tilde{X}^2 + \tilde{X}^3), \, 
[\tilde{X}^2,\tilde{X}^3] = 0, \; 
[\tilde{X}^3,\tilde{X}^1] = b  \tilde{X}^3,$
\\ $b \in { \real} - \{ 0 \} . $ 
   \item[${\bf ( 7_0|5.i) } $] : 
$ [\tilde{X}^1,\tilde{X}^2]=  - \tilde{X}^2 , \, 
[\tilde{X}^2,\tilde{X}^3] = 0, \; 
[\tilde{X}^3,\tilde{X}^1] =  \tilde{X}^3,  $ 
   \item[${\bf ( 7_0|5.ii|b) } $] : 
$  [\tilde{X}^1,\tilde{X}^2]=  0 , \, 
[\tilde{X}^2,\tilde{X}^3] = b \tilde{X}^2, \; 
[\tilde{X}^3,\tilde{X}^1] = -b \tilde{X}^1, 
\, b >0  . $
\end{description}

\vskip2mm  \item Manin triples with the first subalgebra $ \cg= {\bf 6_{a}}$:
\begin{description} 
   \item[${\bf ( 6_{a}|1) } $] : 
$  [\tilde{X}^1,\tilde{X}^2]= 0, \, 
[\tilde{X}^2,\tilde{X}^3] = 0 , \; 
[\tilde{X}^3,\tilde{X}^1] = 0. $ 
   \item[${\bf ( 6_{a}|2) } $] :
$ [\tilde{X}^1,\tilde{X}^2]=0, \, 
[\tilde{X}^2,\tilde{X}^3] = \tilde{X}^1 , \; 
[\tilde{X}^3,\tilde{X}^1] = 0 . $ 
   \item[${\bf ( 6_{a}|6_{1/a}.i|b) } $] : 
$ [\tilde{X}^1,\tilde{X}^2]=- b ( \frac{1}{a} \tilde{X}^2+\tilde{X}^3), \, 
[\tilde{X}^2,\tilde{X}^3] = 0,$ \\ 
$[\tilde{X}^3,\tilde{X}^1] = b (\tilde{X}^2+ \frac{1}{a} \tilde{X}^3), \; b \in { \real} - \{ 0 \} . $ 
   \item[${\bf ( 6_{a}|6_{1/a}.ii) } $] : 
$ [\tilde{X}^1,\tilde{X}^2]= \tilde{X}^1 , \, 
[\tilde{X}^2,\tilde{X}^3] = \frac{a+1}{a-1} (\tilde{X}^2+\tilde{X}^3) , \; 
[\tilde{X}^3,\tilde{X}^1] = \tilde{X}^1 . $
   \item[${\bf ( 6_{a}|6_{1/a}.iii) } $] : 
$ [\tilde{X}^1,\tilde{X}^2]= \tilde{X}^1 , \, 
[\tilde{X}^2,\tilde{X}^3] = \frac{a-1}{a+1} (-\tilde{X}^2+\tilde{X}^3) , \; 
[\tilde{X}^3,\tilde{X}^1] = -\tilde{X}^1 .$ 
\end{description}

\vskip2mm  \item Manin triples with the first subalgebra $ \cg= {\bf 6_0}$:
\begin{description}
   \item[${\bf ( 6_0|1) } $] :
$ [\tilde{X}^1,\tilde{X}^2]= 0, \, 
[\tilde{X}^2,\tilde{X}^3] = 0 , \; 
[\tilde{X}^3,\tilde{X}^1] = 0. $ 
   \item[${\bf ( 6_0|2) } $] :
 $  [\tilde{X}^1,\tilde{X}^2]= \tilde{X}^3, \, 
[\tilde{X}^2,\tilde{X}^3] = 0 , \; 
[\tilde{X}^3,\tilde{X}^1] = 0 .$ 
   \item[${\bf ( 6_0|4.i|b) } $] :
$ [\tilde{X}^1,\tilde{X}^2]= b ( - \tilde{X}^2 + \tilde{X}^3), \, 
[\tilde{X}^2,\tilde{X}^3] = 0,$ \\ 
$[\tilde{X}^3,\tilde{X}^1] = b  \tilde{X}^3, 
\; b \in { \real} - \{ 0 \} . $
   \item[${\bf ( 6_0|4.ii) } $] :
$ [\tilde{X}^1,\tilde{X}^2]=  ( - \tilde{X}^1 + \tilde{X}^2 + \tilde{X}^3), \, 
[\tilde{X}^2,\tilde{X}^3] = \tilde{X}^3, \; 
[\tilde{X}^3,\tilde{X}^1] =  - \tilde{X}^3 
 . $
   \item[${\bf ( 6_0|5.i) } $] : 
$ [\tilde{X}^1,\tilde{X}^2]=  - \tilde{X}^2 , \, 
[\tilde{X}^2,\tilde{X}^3] = 0, \; 
[\tilde{X}^3,\tilde{X}^1] =  \tilde{X}^3. $ 
   \item[${\bf ( 6_0|5.ii) } $] : 
$ [\tilde{X}^1,\tilde{X}^2]=  - \tilde{X}^1+ \tilde{X}^2 , \, 
[\tilde{X}^2,\tilde{X}^3] =  \tilde{X}^3, \; 
[\tilde{X}^3,\tilde{X}^1] = - \tilde{X}^3. $ 
   \item[${\bf ( 6_0|5.iii|b) } $] : 
$ [\tilde{X}^1,\tilde{X}^2]=  0 , \, 
[\tilde{X}^2,\tilde{X}^3] = - b\tilde{X}^2, \; 
[\tilde{X}^3,\tilde{X}^1] =  b\tilde{X}^1, \ b>0 . $ 
\end{description}

\vskip2mm  \item Manin triples with the first subalgebra $ \cg= {\bf 5}$:
\begin{description} 
   \item[${\bf ( 5|1) } $] :
$ [\tilde{X}^1,\tilde{X}^2]= 0, \, 
[\tilde{X}^2,\tilde{X}^3] = 0 , \; 
[\tilde{X}^3,\tilde{X}^1] = 0. $ 
   \item[${\bf ( 5|2.i) } $] :  
$  [\tilde{X}^1,\tilde{X}^2]= 0, \, 
[\tilde{X}^2,\tilde{X}^3] = \tilde{X}^1 , \; 
[\tilde{X}^3,\tilde{X}^1] =  0  .$
   \item[${\bf ( 5|2.ii) } $] : 
$ [\tilde{X}^1,\tilde{X}^2]= \tilde{X}^3, \, 
[\tilde{X}^2,\tilde{X}^3] = 0 , \; 
[\tilde{X}^3,\tilde{X}^1] =  0  .$
\end{description}
  and dual Manin triples ${\bf ( \cg \leftrightarrow \tcg ) } $ to Manin triples given above for  
$\cg = {\bf 6_0}$, ${\bf 7_0} $, ${\bf 8}$, ${\bf 9}$.   

\vskip2mm  \item Manin triples with the first subalgebra $ \cg= {\bf 4}$:
\begin{description} 
   \item[${\bf ( 4|1) } $] :
$ [\tilde{X}^1,\tilde{X}^2]= 0, \, 
[\tilde{X}^2,\tilde{X}^3] = 0 , \; 
[\tilde{X}^3,\tilde{X}^1] = 0. $ 
   \item[${\bf ( 4|2.i) } $] :
 $ [\tilde{X}^1,\tilde{X}^2]= 0, \, 
[\tilde{X}^2,\tilde{X}^3] = \tilde{X}^1 , \; 
[\tilde{X}^3,\tilde{X}^1] =  0  .$
   \item[${\bf ( 4|2.ii) } $] : 
 $ [\tilde{X}^1,\tilde{X}^2]= 0, \, 
[\tilde{X}^2,\tilde{X}^3] = - \tilde{X}^1 , \; 
[\tilde{X}^3,\tilde{X}^1] =  0  .$
   \item[${\bf ( 4|2.iii|b) } $] :
 $ [\tilde{X}^1,\tilde{X}^2]= 0, \, 
[\tilde{X}^2,\tilde{X}^3] = 0 , \; 
[\tilde{X}^3,\tilde{X}^1] = b \tilde{X}^2, \, b \in \real- \{ 0 \} .$ 
\end{description}
  and dual Manin triples ${\bf ( \cg \leftrightarrow \tcg ) } $ to Manin triples given above for 
$\cg = {\bf 6_0}$, ${\bf 7_0}$.  

\vskip2mm  \item Manin triples with the first subalgebra $ \cg= {\bf 3}$:
\begin{description} 
   \item[${\bf ( 3|1) } $] : 
$ [\tilde{X}^1,\tilde{X}^2]= 0, \, 
[\tilde{X}^2,\tilde{X}^3] = 0 , \; 
[\tilde{X}^3,\tilde{X}^1] = 0. $ 
   \item[${\bf ( 3|2) } $] :
$ [\tilde{X}^1,\tilde{X}^2]=0, \, 
[\tilde{X}^2,\tilde{X}^3] = \tilde{X}^1 , \; 
[\tilde{X}^3,\tilde{X}^1] = 0 . $ 
   \item[${\bf ( 3|3.i|b) } $] : 
$ [\tilde{X}^1,\tilde{X}^2]=- b (\tilde{X}^2+\tilde{X}^3), \, 
[\tilde{X}^2,\tilde{X}^3] = 0, \; 
[\tilde{X}^3,\tilde{X}^1] = b (\tilde{X}^2+\tilde{X}^3), \\
 b \in { \real} - \{ 0 \} . $ 
   \item[${\bf ( 3|3.ii) } $] : 
$ [\tilde{X}^1,\tilde{X}^2]= 0 , \, 
[\tilde{X}^2,\tilde{X}^3] = \tilde{X}^2+\tilde{X}^3 , \; 
[\tilde{X}^3,\tilde{X}^1] =  0 . $
   \item[${\bf ( 3|3.iii) } $] : 
$ [\tilde{X}^1,\tilde{X}^2]= \tilde{X}^1 , \, 
[\tilde{X}^2,\tilde{X}^3] = 0 , \; 
[\tilde{X}^3,\tilde{X}^1] = - \tilde{X}^1 . $
\end{description}

\vskip2mm  \item Manin triples with the first subalgebra $ \cg= {\bf 2}$:
\begin{description} 
   \item[${\bf ( 2|1) } $] :
$ [\tilde{X}^1,\tilde{X}^2]= 0, \, 
[\tilde{X}^2,\tilde{X}^3] = 0 , \; 
[\tilde{X}^3,\tilde{X}^1] = 0. $ 
   \item[${\bf ( 2|2.i) } $] : 
 $ [\tilde{X}^1,\tilde{X}^2]= \tilde{X}^3, \, 
[\tilde{X}^2,\tilde{X}^3] = 0 , \; 
[\tilde{X}^3,\tilde{X}^1] =  0  .$
   \item[${\bf ( 2|2.ii) } $] : 
 $ [\tilde{X}^1,\tilde{X}^2]= -\tilde{X}^3, \, 
[\tilde{X}^2,\tilde{X}^3] = 0 , \; 
[\tilde{X}^3,\tilde{X}^1] =  0  .$
\end{description}
 and dual Manin triples ${\bf ( \cg \leftrightarrow \tcg ) } $ to Manin triples given above  for 
$\cg= {\bf 3}$, ${\bf 4}$, ${\bf 6_0}$, ${\bf 6_a}$,  ${\bf 7_0}$, ${\bf 7_a}$.    

\vskip2mm   \item Manin triples with the first subalgebra $ \cg= {\bf 1}$:
\begin{description}
\item[${\bf ( 1|1) } $] : 
$ [\tilde{X}^1,\tilde{X}^2]= 0, \, 
[\tilde{X}^2,\tilde{X}^3] = 0 , \; 
[\tilde{X}^3,\tilde{X}^1] = 0. $ 
\end{description}
and dual Manin triples ${\bf ( \cg \leftrightarrow \tcg ) } $ to Manin triples given above  for 
$\cg= {\bf 2}$--${\bf 9}$.

\end{enumerate}
%\bibliography{pltdualm}

\begin{thebibliography}{10}

\bibitem{klse:dna}
C.~Klim\v{c}\'{\i}k and P.~\v{S}evera, \emph{Dual nonabelian duality
  and the Drinfeld double}, \plb{351}{1995}{455} [\hepth{9502122}].

\bibitem{kli:pltd}
C.~Klim\v{c}\'{\i}k, \emph{Poisson-Lie T-duality},
\npps{46}{1996}{116} [\hepth{9509095}].

\bibitem{hlasno:pltdm2dt}
L.~Hlavat\'y and L.~\v{S}nobl,
\emph{Classification of {P}oisson--{L}ie {T}--dual models with
  two--dimensional targets},
\mpla{17}{2002}{429} [\hepth{0110139}].

\bibitem{klva:olr}
C.Klim\v c\'{\i}k and G.Valent,
\emph{One--loop renormalizability of all 2d dimensional {P}oisson--{L}ie
  sigma models},
\plb{565}{2003}{237} [\hepth{0304053}].

\bibitem{vall:su2}
M.A. Lledo and V.S. Varadarajan, \emph{{\rm SU}(2) Poisson-Lie
T-duality}, \lmp{45}{1998}{247} [\hepth{9803175}].

\bibitem{sfe:pltd}
K.~Sfetsos, \emph{Poisson-Lie T-duality beyond the classical level and
  the renormalization group}, \plb{432}{1998}{365} [\hepth{9803019}].

\bibitem{jare:pltd}
M.A. Jafarizadeh and A.Rezaei-Aghdam,
\emph{Poisson--{L}ie {T}-duality and {B}ianchi type algebras},
\plb{458}{1999}{470} [\hepth{9903152}].

\bibitem{snohla:ddoubles}
L.~\v{S}nobl and L.~Hlavat\'y,
\emph{Classification of six--dimensional real {D}rinfeld doubles},
\ijmpa{17}{2002}{4043} [\Math{QA}{0202210}].

\bibitem{unge:pltp}
R.~von~Unge, \emph{Poisson-Lie T-plurality}, \jhep{07}{2002}{014}
[\hepth{0205245}].

\bibitem{alkltse:qpl}
A.Yu.Alekseev, C.Klim\v c\'{\i}k, and A.A.Tseytlin,
\emph{Quantum Poisson--{L}ie {T}--duality and {W}{Z}{N}{W} model},
\npb{458}{1996}{430} [\hepth{9509123}]


\bibitem{tvu:pltdpi}
E.~Tyurin and R.~von Unge, \emph{Poisson-Lie T-duality: the
  path-integral derivation}, \plb{382}{1996}{233} [\hepth{9512025}].

\bibitem{hlasno:cfn6mt}
L.~Hlavat\'y and L.~\v{S}nobl,
\emph{Classification of 6-dimensional real {M}anin triples},
[\Math{QA}{0202209}].


\bibitem{sno:modsp} L. \v{S}nobl,\emph{On modular spaces of semisimple Drinfeld 
doubles}, \jhep{09}{2002}{018} [\hepth{0204244}]. 

\bibitem{sno:msldd} L. \v{S}nobl, \emph{Modular spaces of low-dimensional Drinfeld
doubles}, in Proceedings of the \emph{23rd Winter School Geometry and
Physics}, Srni, January 2003, \emph{Rend.\ Circ.\ Mat.\ Palermo},
Serie II, Suppl.\ {\bf 72} (2004) 193.

\bibitem{Landau}
L.D. Landau and E.M. Lifshitz, 
\emph{The Classical Theory of Fields},
 Pergamon Press, 1987.
\end{thebibliography}

\end{document}